\documentclass[aps,pra,amsmath,amssymb,superscriptaddress,twocolumn,10pt,longbibliography]{revtex4-2}
\usepackage[utf8]{inputenc}
\usepackage[normalem]{ulem}
\usepackage{amsmath}

\usepackage{times,txfonts}
\usepackage{nicefrac}
\usepackage{xr-hyper}
\usepackage[colorlinks=true,linkcolor=blue,urlcolor=blue,citecolor=magenta,pdfusetitle]{hyperref}
\usepackage{epsfig} 
\usepackage{subfigure}
\usepackage{physics}
\usepackage{bm}
\usepackage{xcolor}
\usepackage{dsfont}

\def\bra#1{\langle{#1}|}
\def\ket#1{|{#1}\rangle}

\begin{document}

\title{Non-Markovian dynamics of generation of bound states in the continuum \\ via single-photon scattering}

\author{Giuseppe Magnifico}
\author{Maria Maffei}\email{maria.maffei@uniba.it}
\author{Domenico Pomarico}
\author{Debmalya~Das}
\author{Paolo Facchi}
\author{Saverio Pascazio}
\author{Francesco V. Pepe}

\affiliation{Dipartimento di Fisica, Universit\`{a} di Bari, I-70126 Bari, Italy}
\affiliation{INFN, Sezione di Bari, I-70125 Bari, Italy}

\begin{abstract}
The excitation of bound states in the continuum (BICs) in two- or multi-qubit systems lies at the heart of entanglement generation and harnessing in Waveguide Quantum Electrodynamics platforms. However, the generation of qubit pair BICs through single-photon scattering is hindered by the fact that these states are effectively decoupled from propagating photons. We prove that scattering of a parity-invariant single photon on a qubit pair, combined with a properly engineered time variation of the qubit detuning, is not only feasible, but also more effective than strategies based on the relaxation of the excited states of the qubits when the distance between the qubits gives rise to non-negligible photon delays (non-Markovian regime). The use of tensor network methods to simulate the proposed scheme enables to include such photon delays in collision models, thus opening the possibility to follow the time evolution of the full quantum system, including qubits and field, and to efficiently implement and characterize the dynamics hence identifying optimal working points for the BIC generation.
\end{abstract}


\maketitle

%\tableofcontents

\section{Introduction}
Waveguide Quantum Electrodynamics (wQED) represents one of the most promising, feasible and versatile platforms for quantum technology implementations~\cite{Ciccarello_OPN2024}. The opportunity of manipulating the system by tuning interaction strengths, emission frequencies, and distances between components, along with the effectiveness of emitter-field and emitter-emitter couplings, guaranteed by quasi-one-dimensional field confinement, open the way to theoretical investigation and technological exploitation of remarkable collective phenomena~\cite{onedim_review, waveguide_njp, cqed1, cqed2, onedim1, onedim2, onedim3, onedim4, onedim5, onedim6, kimble1, focused1, focused2, focused3, mirror1, mirror2, atomrefl1, yudsonPLA,Weng2025}. Among the latter, states with suppressed spontaneous emission \cite{Mahmoodian20,waveguide_pra3,waveguide_njp} are particularly attractive for energy \cite{tirone2024many} and information storage \cite{AsenjoPRX}, and can be implemented in state-of-the-art superconducting circuits~\cite{Cardenas2015}. However, their theoretical description is still in an exploratory stage expecially when the so-called non-Markovian regime is considered~\cite{Pichler2017,maricarmen24,cilluffo2025operator}. Here non-negligible distances between the emitters give rise to photon delays and feedback effects that make the dynamics non-Markovian.

Interesting effects already emerge in the case of a pair of two-level emitters (qubits), coupled to  
one parity-invariant  
waveguide mode in the weak coupling regime~\cite{refereeA1,refereeA2,waveguide_pra,baranger,baranger2013,NJP,yudson2014,laakso,Fedorov1,Fedorov2,Alexia2016,Calajo,waveguide_pra4,Guimond2020unidirectional,gasparinetti,gasparinetti2023photons,waveguide_pra2,maffei2024directional}. When their distance matches resonance conditions, the system Hamiltonian has an eigenstate with energy in the continuous spectrum, i.e. a bound state in the continuum (BIC)~\cite{stillinger1975bound,friedrich1985interfering}. The two-qubit BIC has a single excitation shared among the subradiant Bell state and a single-photon standing wave trapped between them. This suggests a strategy to generate entanglement between qubits from the Hamiltonian evolution: an initial state with one excited qubit and field in vacuum will partially relax towards the BIC~\cite{gullans2012nanoplasmonic,waveguide_pra,tiranov2019}. This procedure, called \textit{entanglement by relaxation}~\cite{waveguide_pra}, is intrinsically limited by the overlap between the initial state and the BIC and thus gives at best a probability one-half of generating the Bell state. Moreover, entanglement by relaxation becomes less and less efficient when the distance between qubits increases and the excitation is more likely stored in the field between them~\cite{waveguide_pra,Calajo,sinha2020collective,sinha2020nonmarkovian}. 

Interestingly, one can harness such stationary field to generate qubits Bell states in an alternative manner. Exciting a BIC through scattering is generally hindered by vanishing overlaps between trapped and incoming photons, thus requiring strategies such as time-dependent Hamiltonians \cite{hayran2021capturing}, and two-photon scattering \cite{Calajo,trivedi2021optimal}. The BIC of distant emitters can also be populated via relaxation of their doubly excited state~\cite{alvarezgiron2023delay}. 

\begin{figure}
    \centering
\includegraphics[width=0.49\textwidth]{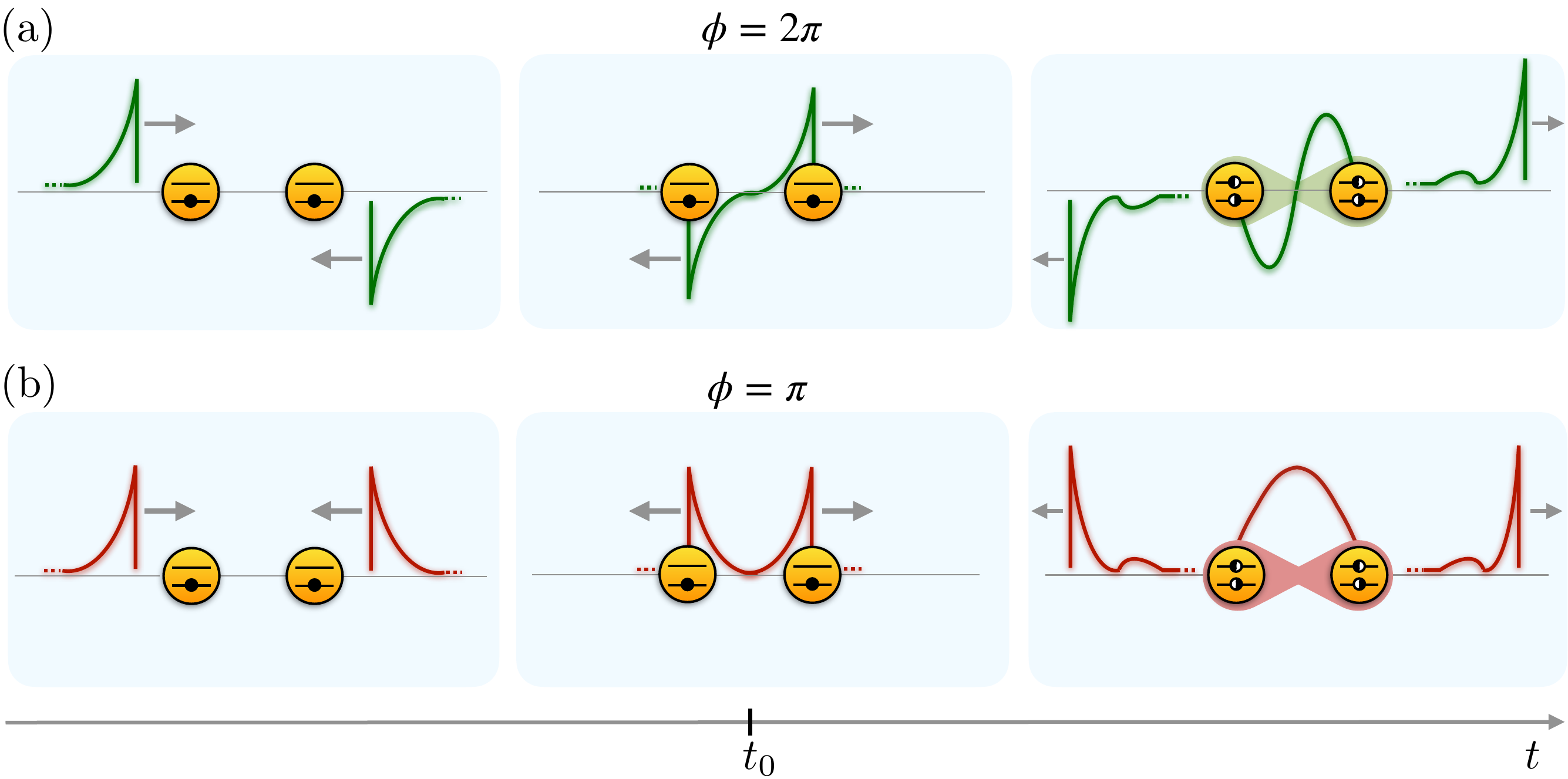} 
\caption{Schematics of BIC generation via single-photon scattering. From left to right, three snapshots at different stages of the proposed protocol. In both panels $\phi = \omega_0\tau$, with $\omega_0$ being the qubits excitation frequency, and $\tau= d/v_g$ being the time that takes to the photon of speed $v_g$ to cover the inter-qubits distance $d$. The system is in an even-order resonance, with $\phi=2\pi$, in panel (a), and in an odd-order resonance, with $\phi=\pi$, in panel (b), with even and odd referring to the BIC spatial symmetry. At $t\ll t_0$, the qubits are in their ground states and decoupled from the field, while the incoming photon is prepared in a spatially antisymmetric [panel (a)] or symmetric [panel (b)] state. Until $t=t_0$, the qubits are detuned with respect to the photon. At $t=t_0$, the wavefronts of right- and left-propagating components of the incoming photon have covered the whole distance between qubits, and the latter are tuned in (detuning is turned off), getting coupled to the field. When the scattering dynamics is completed, at $t\gg t_0$, the qubits-photon system is found, with a finite probability, in a BIC containing a qubits Bell-state (resp. green and red shaded areas) with same spatial symmetry as the incoming photon.}
    \label{fig:scheme}
\end{figure}

In this work, we propose a method to effectively populate a two-qubit BIC in the non-Markovian regime, by sending \textit{one} photon on the qubits in their ground states. The method is depicted in Fig.~\ref{fig:scheme}: in order to bypass the roadblock of the vanishing overlap between the incident photon and the BIC stationary one, we suppress the interaction until, at time $t_0$, the incident photon has traveled the whole distance between the qubits, then we quickly put them back in the dynamics. We specifically propose of quenching the qubit transition frequency from an initially off-resonance value to the resonant one, a task experimentally feasible in the superconducting circuit domain~\cite{gasparinetti_switch,wilson2011observation,WilsonPRL2010}. We will show that this strategy, called \textit{quenched detuning}, is effective in populating the BIC. To account for non-perturbative effects in the system dynamics in the non-Markovian regime, we employ tensor network techniques, specifically Matrix Product States (MPS), providing accurate numerical solutions in the time domain~\cite{ZollerCM,Pichler2017,guimond2017scattering, cilluffoMPO, vodenkova2023continuous, papaefstathiou2024efficienttensornetworksimulation}.

\section{Model and dynamics}
We consider a pair of identical qubits, with excitation frequency $\omega_0$, coupled to a waveguide mode at points $x_1=0$ and $x_2$, respectively, with bare Hamiltonians
\begin{equation}
    H^{(0)}_j= \hbar \omega_0 \sigma^{\dagger}_j\sigma_j^{\,}, \quad \text{with } \sigma_j=|g_j\rangle\langle e_j| ,
\end{equation}
with $j\in\{1,2\}$ and $\ket{g_j}$ and $\ket{e_j}$ being ground and excited state, respectively. For notation shortness, we will denote the tensor product basis of the qubit pair as $\ket{ee}:=\ket{e_1}\otimes\ket{e_2}$, $\ket{eg}:=\ket{e_1}\otimes\ket{g_2}$, $\ket{ge}:=\ket{g_1}\otimes\ket{e_2}$, and $\ket{gg}:=\ket{g_1}\otimes\ket{g_2}$. 
Following the circuit-QED paradigm~\cite{Gardiner1985Input}, we consider the field perfectly confined in one dimension. Moreover, we assume that its dispersion relation can be linearized, with group velocity $v_g$, in the frequency range of interest for the interaction. It is hence convenient to define the wavenumber $k_0=\omega_0/v_g$, corresponding to the qubits transition frequency. We account for photon propagation in two possible directions by introducing the field operators $a_{R(L)}(\omega)$, that annihilate right(left)-propagating photons of frequency $\omega$ and wavenumber $k=+(-)\omega/v_g$ and satisfy the canonical commutation relations $[a^{\,}_{D}(\omega),a^{\dagger}_{D'}(\omega')]=\delta_{D,D'}\delta(\omega-\omega')$ and $[a_{D}(\omega),a_{D'}(\omega')] = 0$, with $D,D'\in\lbrace R,L\rbrace$. The bare field Hamiltonian hence reads
\begin{equation}
    H^{(0)}_{\text{f}}=\hbar \sum_{D=R,L}\int d\omega~\omega \ a^{\dagger}_{D}(\omega)a^{\ }_{D}(\omega) .
\end{equation}
In the interaction picture of $H^{(0)} = H^{(0)}_1 + H^{(0)}_2 + H^{(0)}_{\text{f}}$, the emitters-field coupling Hamiltonian reads (see Appendix~\ref{Appendix_sec_1})
\begin{align}\label{eq:V_continuousTime}
    V_{I}(t)=& \hbar \sqrt{\frac{\gamma}{2}}\Biggl\lbrace \sigma_{1} \left[ b^{\dagger}_{R}\left(t\right)+  b^{\dagger}_{L}\left(t\right)\right]   \\ \nonumber 
    & +\sigma_{2} \left[e^{-i \phi} b^{\dagger}_{R}\left(t-\tau\right)+ e^{i \phi} b^{\dagger}_{L}\left(t+\tau\right)\right] \Biggr\rbrace + \mathrm{H.c.} ,
\end{align}
where $\tau=d/v_g$ is the photon propagation time between the qubits, $\phi=k_0 d= \omega_0 \tau$ is the corresponding phase, and $b_{D}(t)= (2\pi)^{-\frac{1}{2}}\int d\omega\, a_{D}(\omega)e^{-i(\omega-\omega_0)t}$ are quantum noise operators~\cite{gardinerzoller} satisfying $[b_{D}(t),b^{\dagger}_{D'}(t')]=\delta_{D,D'}\delta(t-t')$ with $D,D'\in\lbrace R,L\rbrace$. Equation~\eqref{eq:V_continuousTime} holds for weak qubit-field coupling that leads to the validity of 
the rotating wave approximation and the assumption of a flat coupling over the relevant bandwidth~\cite{cohentannoudjiAPI,burgarth2023taming}. When the propagation delay $\tau$ between the qubits is much smaller than $\gamma^{-1}$, the effect of the finite propagation time can be neglected and the qubits emission dynamics can be modeled with canonical Born-Markov master equations   ~\cite{combes_slh_2017,GKS,L}. The detuning Hamiltonian
\begin{equation}
\label{eq:detuning}
    H_{\mathrm{det}}(t) = \hbar \Delta\omega (\sigma^{\dagger}_{1}\sigma_{1}+ \sigma^{\dagger}_{2}\sigma_{2}) \, \Theta(t_0-t)
\end{equation}
with $\Theta(t)$ the Heaviside step function, is turned off at $t=t_0$, when the incoming photon crosses the whole inter-qubit distance (see Fig.~\ref{fig:scheme}).

\begin{figure*}
    \centering
    \includegraphics[width=0.99\textwidth]{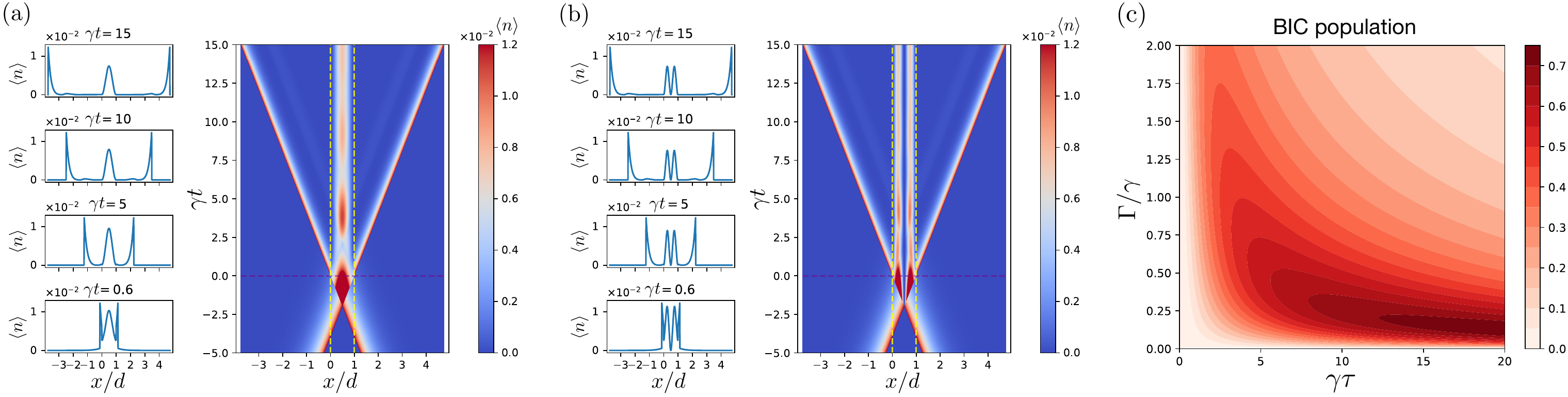}
    \caption{Dynamics of the BIC generation via single-photon scattering. All the data are obtained by MPS numerical simulation on a lattice of $10^3$ collision units, setting $\gamma \Delta t = 4 \times 10^{-2}$, and truncation error in the state vector norm at the end of the time evolution equal to $10^{-4}$. (a--b) Time-evolution of the average photon number per site $\langle n(x)$, for a delay $\tau=4\gamma^{-1}$ (with $\gamma$ the excited level lifetime), for $\phi=\pi$ and initial state $\ket{\Psi(0)}=\ket{gg} \otimes \ket{\varphi_{+}}$ (a), and $\phi=2\pi$ and initial state $\ket{\Psi(0)}=\ket{gg}\otimes \ket{\varphi_{-}}$ (b). The light-matter coupling is switched on at time $t_0=0$ when the right-propagating wave-front hits $x=d$ and the left-propagating one hits $x=0$ (see Fig.~\ref{fig:scheme}). The incoming wavepackets have bandwidth $\Gamma = 0.625\,\gamma$. The insets show $\langle n(x)\rangle$ at different times during the system's evolution. (c) Asymptotic probability (final time) of populating the BIC starting from initial state $\ket{\Psi(0)}=\ket{gg} \otimes \ket{\varphi_{(-)^{\ell +1}}}$ varying the incoming wavepacket's bandwidth and the inter-qubits delay $\tau$ of integer multiples, $\ell$, of $\pi/\omega_0=\Delta t$ such that the system is always at resonance. The plot obtained from the simulation data perfectly matches the analytical result of Eq.~\eqref{eq:th_upp_bound}.} 
    \label{fig:dynamics}
\end{figure*}

We set a coarse-graining time interval $\Delta t\ll\gamma^{-1}$ and introduce the quantum noise increment operators, $B_{n,D}\equiv (\Delta t)^{-1/2}\int_{n \Delta t}^{(n+1)\Delta t} \mathrm{d}t'~b_{D}(t')$, with $n$ being an integer, 
annihilating an excitation of the $n-$th discrete-time field mode, or \textit{collision unit}, in the direction $D$ \cite{gardinerzoller, Ciccarello_2017,Ciccarello2020,CICCARELLO2021}. The coarse-grained evolution of the system is given by the unitary map
\begin{equation}\label{eq:collision_Un}
    \ket{\Psi(t_{n+1})} = \exp\Bigl \lbrace-i \left[ O_{1}(t_n)+O_{2}(t_n) + H_{\mathrm{det}}(t_n) \Delta t/\hbar \right] \Bigr\rbrace \ket{\Psi(t_{n})}
\end{equation}
with
\begin{align} \label{eq:longrange}
O_{1}(t_n)&=\sqrt{\frac{\gamma \Delta t}{2}} \sigma_{1} \left[ B^{\dagger}_{n,R}+  B^{\dagger}_{n,L}\right]+ \mathrm{H.c.} , \\ \nonumber
O_{2}(t_n)&=\sqrt{\frac{\gamma \Delta t}{2}} \sigma_{2} \left[e^{-i \phi} B^{\dagger}_{n-\ell,R}+ e^{i \phi} B^{\dagger}_{n+\ell, L}\right]+ \mathrm{H.c.} ,
\end{align}
where the integer $\ell=\tau/\Delta t= d/(v_g \Delta t)$ is fixed by the distance $d$ between the qubits. The map~\eqref{eq:collision_Un} couples the qubits with different collision units at any (discrete) time of the evolution, its structure features the memory effects due to propagation delay: the $n$-th right-propagating collision unit that interacts with qubit 1 at $t=t_n$, will interact with qubit 2 at $t=t_n+\tau=t_{n+\ell}$; similarly the $n$-th left-propagating unit that interacts with qubit 1 at $t=t_n$ has already interacted with qubit 2 at $t=t_n-\tau=t_{n-\ell}$.

This non-Markovian dynamics can be efficiently simulated by means of a MPS simulation, as first proposed in Ref.~\cite{ZollerCM}. In this framework, a specific MPS site hosts both the qubits, whereas all the other sites encode the left- and right-propagating collision units. The implemented algorithm drives the evolution of the joint qubits-field system state once  the initial state is initialized in the MPS form. This evolution is carried out by iteratively applying to the MPS the quantum gates forming the unitary map in Eq.~\eqref{eq:collision_Un}. Technical details on the algorithm and the initial state preparation are given in Appendix~\ref{Appendix_sec_2}.
 
\section{Results}
It is convenient to exploit the linearity of the dispersion relations and introduce the canonical spatial mode operators
\begin{equation}
    c_R(x) = \frac{1}{\sqrt{v_g}} e^{i k_0 x} b_{R}(-x/v_g), \quad c_L(x) = \frac{1}{\sqrt{v_g}} e^{-i k_0 x} b_{L}(x/v_g) , 
\end{equation}
that allow to express the BIC eigenstates \cite{waveguide_pra,Calajo,sinha2020collective,sinha2020nonmarkovian} at resonance, $\phi= n \pi$, as
\begin{eqnarray}
    \ket{\Phi_{\pm}} &=& \left( 1 + \frac{\gamma\tau}{2} \right)^{-\frac{1}{2}}  \biggl [ \ket{\psi_{\pm}} \otimes \ket{0} 
    + \sqrt{\frac{\gamma}{2 v_g}} \ket{gg} \otimes \ket{1_{\mathrm{BIC}}}\biggr],
 \nonumber\\
   \ket{1_{\mathrm{BIC}}}&=&     \int_{0}^{d} dx \frac{ e^{i k_0 x} c^{\dagger}_{R}(x)-e^{-i k_0 x} c^{\dagger}_{L}(x) }{\sqrt{2} i}\ket{0},
   \label{eq:BIC_m}
\end{eqnarray}
where $\ket{0}$ is the field vacuum, $\ket{\psi_{\pm}}=(\ket{eg}\pm\ket{ge})/\sqrt{2}$ are Bell states of the qubit pair, and $\pm=(-1)^{n+1}$. Equation~\eqref{eq:BIC_m} shows that the photonic part of the BIC, $\ket{1_{\mathrm{BIC}}}$, corresponds to a stationary field trapped between the qubits, with 
local energy density  
$\bra{1_{\mathrm{BIC}}}c^\dagger(x)c(x)\ket{1_{\mathrm{BIC}}}= 2 \sin^2{(k_0 x)}$, where $c(x) = c_{R}(x)+c_{L}(x)$.

In order to effectively populate the BIC starting from the qubits in $\ket{gg}$ we apply the scattering method depicted in Fig.~\ref{fig:scheme}. We prepare the field in a one-photon state having the same spatial symmetry as the photonic part of the BIC, i.e. $\ket{\varphi_{\pm}}=\left(\ket{\varphi_R}\pm \ket{\varphi_{L}}\right)/\sqrt{2}$ where $\ket{\varphi_{D}}=\int dx \, \varphi_{D}(x) \, c^{\dagger}_{D}(x)\ket{0}$ is a single-photon wavepacket of arbitrary shape, propagating in the direction $D$ (see Appendix~\ref{Appendix_sec_3}). 
\begin{figure*}
    \centering
\includegraphics[width=0.99\textwidth]{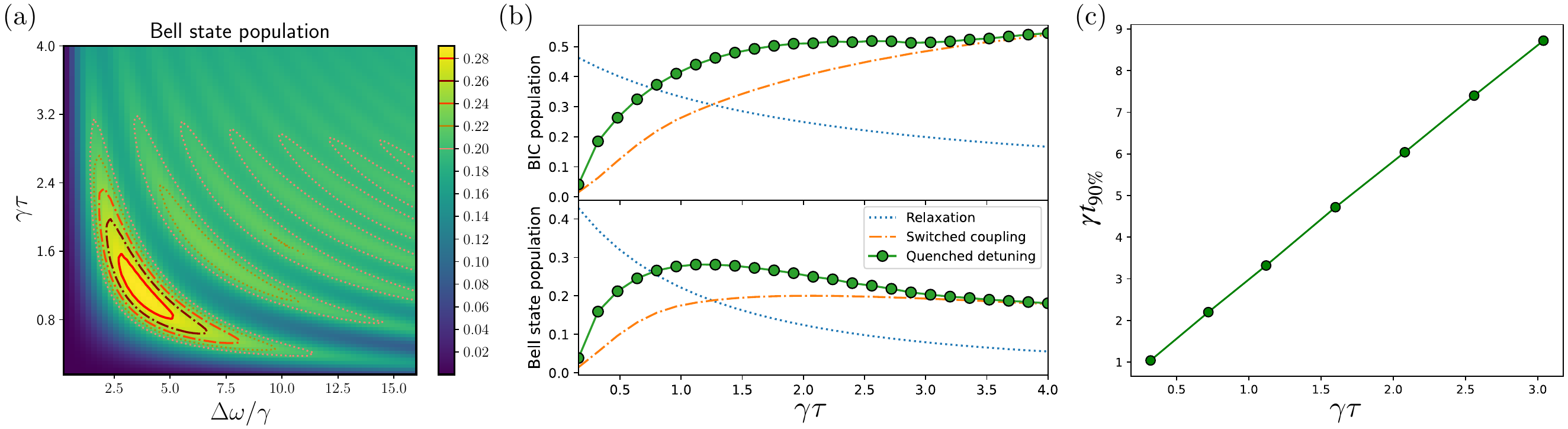}
    \caption{Generation of the BIC via single photon scattering and quenched detuning. (a) Asymptotic (final-time) probability of populating the qubits Bell state starting from an initial state $\ket{\Psi(0)}=\ket{gg}\otimes\ket{\varphi_{(-)^{\ell +1}}}$, for different inter-qubits delays $\tau=\ell \pi/\omega_0$, with $\ell$ being an integer, and different values of the detuning. The value of $\Gamma\tau$ is fixed at $2.51$, which maximizes the BIC population in Eq.~\eqref{eq:th_upp_bound} at infinite detuning. (b) Maximal asymptotic probabilities of populating the BIC and the qubits Bell states versus the inter-qubits delay (green dots). Dotted blue lines correspond to entanglement by relaxation, dashed orange lines to single-photon scattering with infinite detuning (Eq.~\eqref{eq:th_upp_bound}). (c) Time needed to achieve the $90\%$ of the asymptotic BIC population with the quenched detuning method. 
    Each point was computed by using the wavepackets bandwidth and the detuning that maximizes the Bell state population for that value of $\tau$. }
    \label{fig:populations}
\end{figure*}
At time $t_0=0$, the right(left)-propagating wavepackets reach the position of qubit 2(1), and the qubits-field interaction is switched on. In this configuration, the probability \eqref{eq:BIC_m} of populating the BIC coincides with the squared overlap of the latter with the initial state $\ket{gg}\otimes\ket{\varphi_{\pm}}$. In the rest of the paper, we will mostly consider exponentially-shaped wavepackets, naturally produced by qubit decay in one dimension, $\varphi_{R}(x)=\sqrt{\Gamma/v_g} \ \Theta(d-x) \ \exp\left \lbrace -\Gamma (d-x)/(2v_g) +i k_0 (x-d)\right\rbrace$ and $\varphi_{L}(x)=\sqrt{\Gamma/v_g} \ \Theta(x)  \exp\left \lbrace-\Gamma x/(2v_g) -i k_0 x\right\rbrace$, and use the bandwidth $\Gamma$ as an optimization parameter. 

We start by considering the ideal case of infinite detuning, in which the qubits are entirely screened from the field for $t<t_0$. Figure~\ref{fig:dynamics}(a-b) show the results of the MPS simulation of such scattering dynamics, obtained over a lattice of $10^3$ collision units, by setting the collisional coarse-graining time $\Delta t$ equal to $4\times 10^{-2}\gamma^{-1}$, and the truncation error in the state vector norm at the end of the time evolution equal to $10^{-4}$. The plots show the evolution of the average photons number at each point of the $x$ axis, i.e. 
$\langle n(x)\rangle\equiv \langle c^{\dagger}(x) c(x)\rangle$ obtained for an odd and even resonance respectively. The data show the process of standing-wave buildup between the qubits while reaching the long-time limit where it assumes a clear sinusoidal profile (see the insets), as expected from Eq.~\eqref{eq:BIC_m}. The transmitted photon amplitude forms the two main propagating wavefronts, followed with delay~$\tau$ by less intense wavefronts due to the partial reflection of the incoming field. These reflected wavepackets travel towards the edges with a retardation from the main wavefronts equal to the delay~$\tau$ and in the long-time limit (inset) form the two small bumps picked at distance $d$ from each edge.

The probability of populating the BIC with the exponential wavepacket in input reads:
\begin{align}\label{eq:th_upp_bound}
P_{\mathrm{BIC}}=\left \vert \bra {\Phi_{\pm}} \Bigl(\ket{gg} \otimes \ket{\varphi_{\pm}}\Bigr) \right \vert^2=\frac{2\gamma  \left(1-e^{- \Gamma \tau /2 }\right)^2}{\Gamma \left(1+\frac{\gamma\tau}{2}\right) },
\end{align}
which in turn gives the qubit Bell-state population by $P_{\mathrm{Bell}}= \left(1+\gamma\tau/2 \right)^{-1}P_{\mathrm{BIC}}$. The BIC excitation probability is represented as a function of the dimensionless parameters $\gamma\tau$ and $\Gamma/\gamma$ in Fig.~\ref{fig:dynamics}(c). As expected, it grows with the weight of the photonic part of the BIC, i.e. when the delay $\tau$ increases with respect to the excited-state lifetime $\gamma^{-1}$, provided that the duration $\Gamma^{-1}$ of the input pulses also increases with respect to it (i.e. the ratio $\Gamma/\gamma$ decreases). The probabilities are maximized by $\Gamma=\Gamma^* = 2.51/\tau$, corresponding to 
\begin{equation}
    P_{\mathrm{BIC}}^* = 0.41 \frac{\gamma\tau}{1+\frac{\gamma\tau}{2}}, \quad P_{\mathrm{Bell}}^* = 0.41 \frac{\gamma\tau}{\left(1+\frac{\gamma\tau}{2}\right)^2} .
\end{equation}
Remarkably, even though the considered optimization is performed for a particular wavepacket shape, $P^{*}_{\mathrm{BIC}}$ is not far from the absolute upper bound, i.e. the weight of the photon component of the BIC [Eq.~\eqref{eq:BIC_m}], that is reached when the input wavepacket $\varphi_{D}$ is shaped as a square-wave of length $d$ (see Appendix~\ref{Appendix_sec_3} for further details). Notice that in this case, the BIC population can reach 100\% for large inter-qubit distances, as in the two-photon protocol proposed in Ref.~\cite{trivedi2021optimal}.  It is finally worth noting that applying the proposed protocol with a photon wavepacket incoming from one direction only (equivalently $\ket{\varphi_R}$ or $\ket{\varphi_L}$) provides a BIC excitation probability of $P_{\mathrm{BIC}}/2$. Thus, despite such a one-side photon protocol being evidently suboptimal in terms of generation probability, it is still more effective that relaxation from an excited qubit state in a certain range of $\gamma\tau$, and can be useful whenever implementing the overlap of $\ket{\varphi_R}$ and $\ket{\varphi_L}$ turns out to be unfeasible.

We now describe the relevant results obtained for a \textit{finite} detuning $\Delta\omega$, providing at the same time a more realistic physical description and a further optimization parameter. Figure~\ref{fig:populations}(a) shows the probability of populating the qubit Bell state through a finite quenched detuning for each value of $\gamma\tau$, by fixing the wavepacket bandwidth to the value $\Gamma^*$, which maximizes Eq.~\eqref{eq:th_upp_bound}.  Figure~\ref{fig:populations}(b) shows the maximal probability of BIC and Bell state generation with optimal quenched detuning (green dots) compared to the other protocols mentioned so far. The comparison of the three methods clearly shows that in the case $\gamma\tau\lesssim 1$, which includes the Markovian regime, entanglement by relaxation provides the higher probability to excite both the BIC and the qubit Bell state. For larger qubit distances, the proposed scattering approach tends to outperform the others. In general, optimal quenched detuning allows for an improvement with respect to the ideal switch case (infinite detuning). As one can observe from Fig.~\ref{fig:populations}(a), this is due to the fact that the BIC generation probability oscillates with $\Delta\omega$ around the infinite-detuning limit. 
 From a physical point of view, the detuned Hamiltonian [Eq.~\eqref{eq:detuning}] prepares during the time $[t_0-\tau,t_0]$ a more or less optimized initial state for the evolution under $H^{(0)}+V_I$ at $t>t_0$. The increase in performance for quenched detuning remains relevant up to $\gamma\tau\lesssim 3$, where its performance becomes very close to the ideal switch case.
 While the BIC excitation probability increases with the qubits distance, the probability to generate the corresponding qubits Bell state (either singlet or triplet, according to the BIC's symmetry) reaches a maximum at a finite distance. This behavior is due to the decreasing share of qubit excitation in the BIC, counterbalancing the increase in the excitation probability of the latter.  Figure~\ref{fig:populations}(c) displays that the time required to achieve $90\%$ of the asymptotic value of the Bell state population is, with very good approximation, proportional to the distance between the qubits. 

\section{Conclusions}
We proposed a scattering method to populate the one-excitation BIC that appears in the spectrum of a system of two qubits coupled to the same waveguide mode, by exploiting only the single-excitation sector. The method consists in sending on the qubits in their ground states a photon with counterpropagating components, characterized by the same symmetry as the target BIC. We first investigate the case in which the interaction between qubits and field is switched on when the opposite wavefronts have covered the whole inter-qubits distance. With this protocol, we find an improvement with respect to entanglement generation by relaxation, another protocol operating in the one-excitation subspace. The same procedure can be performed, in a suboptimal but simpler way, by sending the photon only from one side, at the price of losing half of the BIC generation probability. The experimental realization of the proposed scheme is subject to the detuning quench time being negligible time with respect to $\gamma^{-1}$ and $\tau=n\pi/\omega_0$. A comparison of the relevant qubit timescales described in Ref.~\cite{hoi2015probing} (lifetime larger than 10 ns, delay time for $n=1$ around 1 ns) to the state-of-the-art bandwidth, larger than 10 GHz, of impedance switches by the use of magnetic fluxes \cite{wilson2011observation} corroborates the experimental feasibility of the described protocol.

Our results show that, at inter-qubit distances $\tau>\gamma^{-1}$, where the system dynamics can be considered non-Markovian, the proposed method overcomes the upper limits of generation by relaxation in the one-excitation \cite{waveguide_pra} and in the two-excitation sectors \cite{alvarezgiron2023delay}. Being rooted in the overlap between the input photon and the photonic part of the BIC, the effectiveness of the proposed method in populating the BIC increases with the distance between qubits, as the share of field excitation in the BIC becomes larger. Interestingly, even with a one-side photon, this method becomes more effective in populating the BIC than entanglement by relaxation, in the limit of very long delays. Then, simulating the system dynamics with a MPS evolution, we characterized the performance of the process in a more realistic setting, in which the effective qubit-field interaction is controlled by a finite-frequency detuning. Surprisingly, due to phase effects, the detuning can be adjusted to even outperform the result achieved in the ideal switch case, especially in the range $\tau < 3\gamma^{-1}$, thus further improving the effectiveness of the method.

The results obtained in this work represent the basis for a more efficient generation process of both BICs and their Bell-state components, e.g. by performing adiabatic elimination of the field.
These pathways will be the subject of future research.

\section{acknowledgments}
We acknowledge support from INFN through the project ``QUANTUM'' and from the Italian funding within the ``Budget MUR - Dipartimenti di Eccellenza 2023--2027''  - Quantum Sensing and Modelling for One-Health (QuaSiModO). PF acknowledges support from the Italian National Group of Mathematical Physics (GNFM-INdAM) and from PNRR MUR project CN00000013-``Italian National Centre on HPC, Big Data and Quantum Computing''. GM acknowledges support from the University of Bari via the 2023-UNBACLE-0244025 grant. DD, MM, SP and FVP acknowledge support from PNRR MUR project PE0000023-NQSTI.

\appendix

\section{Coarse-grained unitary evolution from Hamiltonian description}\label{Appendix_sec_1}

In Schr\"{o}dinger picture, the interaction Hamiltonian of a chain of $N$ identical emitters (placed at equal distances $d$) in the waveguide field reads
\begin{align}
V=\hbar\sum_{j=1}^{N}\int d\omega \left[\sqrt{\frac{\gamma_{j,R}}{2\pi}}e^{i\omega(j-1)\tau}a_{R}(\omega) + \right. \\\nonumber \left. +\sqrt{\frac{\gamma_{j,L}}{2\pi}} e^{-i\omega(j-1) \tau}a_{L}(\omega)\right]\sigma^{\dagger}_{j} + \mathrm{H.c.} 
\end{align}
with $\tau=d/v_g$.
In the interaction picture with respect to $H^{(0)} = \hbar \omega_0 \sum_{j=1}^{N} H^{(0)}_{j}+ H^{(0)}_{\text{f}}$, it reads
\begin{align}
    V_{I}(t)=\hbar \sum_{j=1}^{N}  \left[\sqrt{\gamma_R}e^{i (j-1)\omega_0 \tau} b_{R}\left(t-(j-1)\tau\right)+ \right. \\ \nonumber
    \left. +\sqrt{\gamma_L} e^{-i (j-1)\omega_0\tau} b_{L}\left(t+(j-1)\tau\right)\right] \sigma^{\dagger}_{j} + \mathrm{H.c.} ,
\end{align}
with $b_{R/L}(t)$ quantum noise operators. For $N=2$, $\gamma_R=\gamma_L$, $\omega_0\tau=\phi$, the above Hamiltonian gives Eq.~\eqref{eq:V_continuousTime}.

It can be convenient to shift the left-propagating quantum noise operators:
\begin{align}
b_{L}(t)\rightarrow e^{i \phi}b_{L}(t-\tau),
\end{align}
then one gets the equivalent interaction Hamiltonian
\begin{align} \label{eq:ops_O_2atoms}
\tilde{V}_{I}(t)=\hbar \sqrt{\frac{\gamma}{2}} \left\lbrace \sigma^{\dagger}_{1} \left[ b_{R}(t)+ e^{i\phi}  b_{L}(t-\tau)\right]
+ \right. \\ \nonumber
\left.+\sigma^{\dagger}_{2} \left[e^{i \phi} b_{R}(t-\tau)+ b_{ L}(t)\right]\right\rbrace+ \mathrm{H.c.} 
\end{align}
When we move to the discrete-time description, and coarse-grain the evolution, the interaction $\tilde{V}_{I}(t)$ gives rise to the unitary evolution operator equivalent to the one in Eqs.~\eqref{eq:collision_Un}-\eqref{eq:longrange}, but involving only time-bins number $n$ and $n-\ell$, hence providing an easier MPS mapping:  
\begin{align}\label{eq_unitary2}
\tilde{U}_{n}=\exp\left \lbrace -i\sqrt{\frac{\gamma \Delta t}{2}} \left[\sigma_{1} \left( B^{\dagger}_{n,R}+  e^{-i\phi}B^{\dagger}_{n-\ell,L}\right)+\right.\right. \\ \nonumber
\left.\left.+\sigma_{2} \left(e^{-i \phi} B^{\dagger}_{n-\ell,R}+  B^{\dagger}_{n, L}\right)+ \mathrm{H.c.}\right]\right\rbrace.
\end{align}

Equation~\eqref{eq_unitary2} features long-range couplings between the emitters and the field time-bins that make the dynamics non-Markovian. When $\tau\gamma\ll1$, the delay $\tau$ is within the coarse-graining time $\Delta t$. This in turn implies that $B_{n}\approx B_{n-\ell}$. In this limit Eq.~\eqref{eq_unitary2} features only local interactions and thus, when the initial state of the field does not have correlations between the time-bins, it is always possible to derive a Master Equation for the emitters. 

\begin{figure}
\centering
\includegraphics[width=0.90\linewidth]{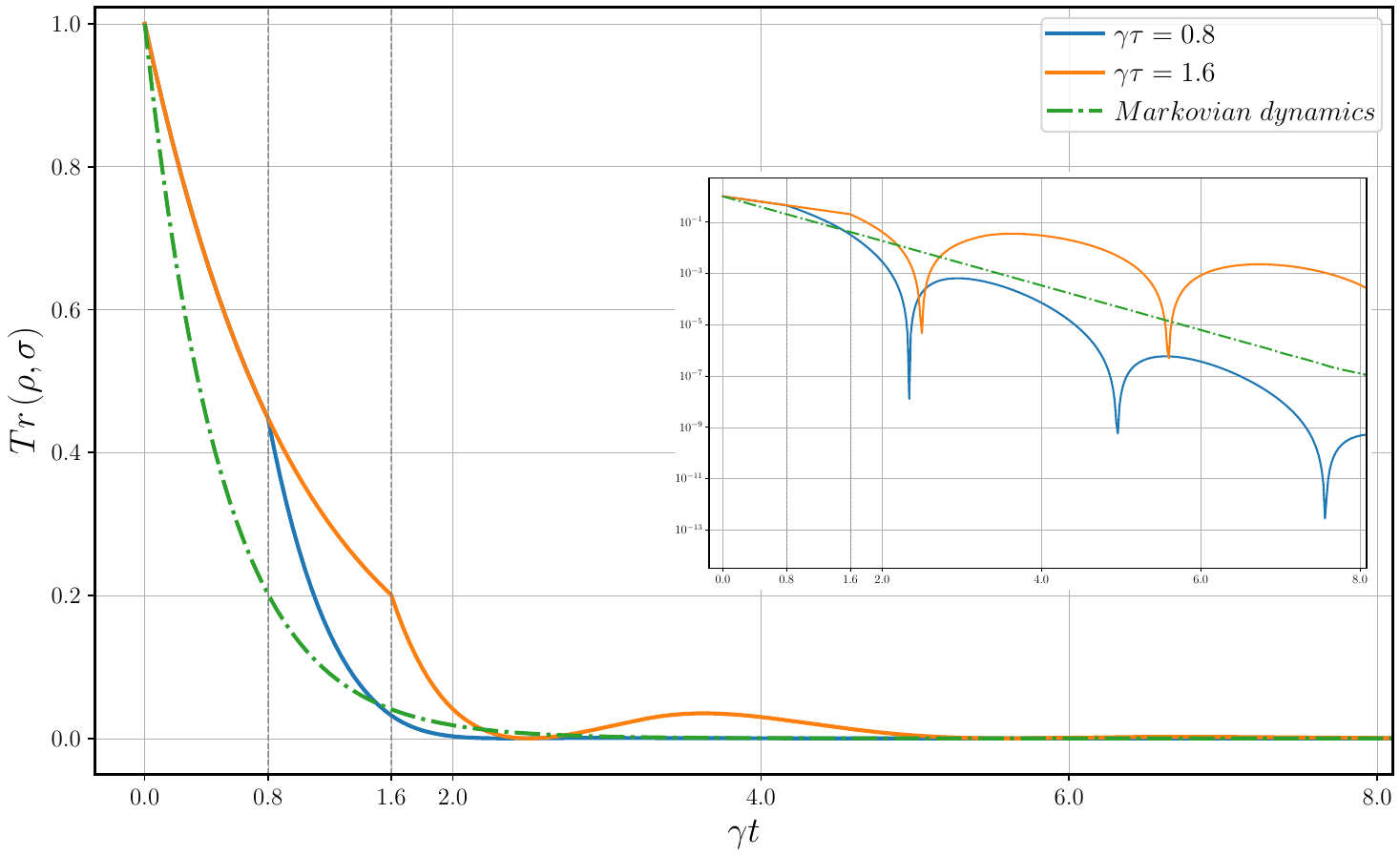}
\caption{Spontaneous emission profiles of the qubits starting from $\rho(0)=\ket{\psi_{+}}\bra{\psi_{+}}$ for different values of the adimensional delay parameter $\gamma\tau=\ell \Delta t$, by keeping the system in an even resonance $\phi=2 \ell\pi$. The curve shows the trace distance between the qubits state at time $t$, $\rho$, and their ground state, $\sigma=\ket{gg}\bra{gg}$, for $\gamma\tau \rightarrow 0$ (Markovian regime) it converges to the exponential $p_{e}(t)=e^{-2\gamma t}$. Points where the first derivative becomes positive signal non-Markovianity.}
\label{fig:decay_comparison}
\end{figure}

In Figure \ref{fig:decay_comparison} we compare the spontaneous emission profiles of the qubits pair for different values of the adimensional delay parameter $\gamma\tau=\ell \Delta t$, by keeping the system in an even resonance $\phi=2 \ell\pi$. Initially the field is in the vacuum state and the qubits are in the radiant Bell state  $\ket{\psi_{+}}$. The decay profile at time $t$ is the trace distance of the qubits state at that time from their ground state. In a Markovian evolution, the derivative of the trace distance is less than or equal to zero for any initial state of the qubits: violation of that property is a signature of non-Markovianity~\cite{Breuer2009}. The dashed line represents the limit of zero-delay where the decay profile is  exponential $p_{e}(t)=e^{-2\gamma t}$, i.e. Markovian evolution. Solid lines correspond to $\gamma\tau= 0.8$ and  $\gamma\tau=1.6$, they show a discontinuity at $\gamma t= \gamma \tau$, and points with positive derivative, signature of non-Markovianity. For increasing values of the time delay, there is sharper non-monotonic behavior, signaling the increased time required by the system to completely emit the field. The latter moves along the delay during the radiant state non-Markovian dynamics, partially increasing its population before the completed emission.

\section{Tensor Network methods}\label{Appendix_sec_2}

\begin{figure*}
    \centering
\includegraphics[width=0.7\textwidth ]{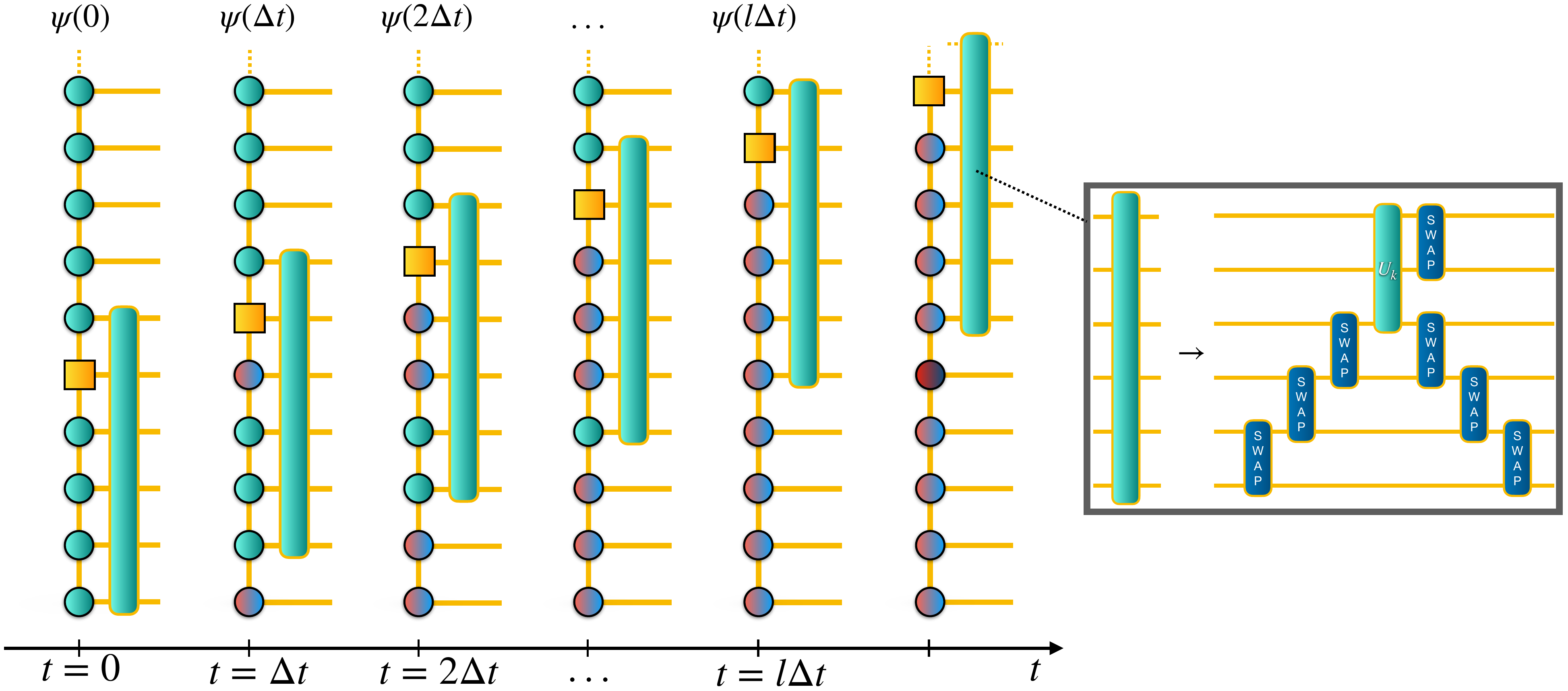}
    \caption{Tensor Network implementation of the system dynamics.}
    \label{fig:MPS_scheme}
\end{figure*}

The MPS scheme for non-Markovian collision models is based on the iterated application of the unitary map in Eq. \eqref{eq_unitary2}, depicted in Figure~\ref{fig:MPS_scheme} \cite{ZollerCM}. The most difficult operation for the application of collision model techniques is referred to the long-range coupling, targeted in tensor network methods by means of $SWAP$ gates as explained in the inset of Figure~\ref{fig:MPS_scheme}.

The time evolution simulation is based on the definition of a time-bins lattice, conveniently modeled in the considered single-excitation sector as a qubit pair per site. Such qubits are not referred in this framework to the two-level emitters undergoing the possible formation of a BIC, but just as a truncated way to host bidirectional photons propagating through the waveguide. Let us consider a unidirectional (or equivalently known as chiral) waveguide, such that each time bin hosting photons can be modeled as a single qubit. If we assume that the propagation is towards the left ($L$), we can express an input photon $\ket{\xi_L} = \Xi \ket{0}$ in terms of a Matrix Product Operator (MPO)
\begin{equation}
    \Xi = W^{[1]} W^{[2]} \dots W^{[N]},
\end{equation}
defined in an analogous way in~\cite{cilluffoMPO}, where $N$ is the number of sites composing the lattice and each operator-valued matrix reads
\begin{align}
    W^{[1]} = \begin{pmatrix}
        \xi(1) \sigma^+_{L,1} & \mathds{1}_2
    \end{pmatrix}, W^{[s]} = \begin{pmatrix}
        \mathds{1}_2 & 0 \\
        \xi(s) \sigma^+_{L,s} & \mathds{1}_2
    \end{pmatrix} \\ \nonumber
    \mathrm{for} \ s=2,\dots,N-1, W^{[N]} = \begin{pmatrix}
        \mathds{1}_2 \\
        \xi(N) \sigma^+_{L,N}
    \end{pmatrix},
\end{align}
yielding $\ket{\xi_L} = \sum_{s} \xi_L(s) \sigma^+_{L,s} \ket{0}$, with $\mathds{1}_2$ standing for a $2 \times 2$ identity matrix. The wave packet is then propagated by applying the evolution represented in Figure~\ref{fig:MPS_scheme}, where circle sites host photonic amplitudes, while the yellow square site is referred to the two-level emitters pair, also implemented as a qubit pair like in the aforementioned bidirectional case. Within each time step a sequence of $SWAP$ gates brings the photonic site placed at the time delay distance $\ell$ at the nearest-neighbor position with respect to the yellow square site, then the unitary map in Eq.~\eqref{eq_unitary2} acts on three contiguous lattice sites. After the interaction another sequence of $SWAP$ gates brings back the photonic site at the time delay position, while a single $SWAP$ gate moves the emitters yellow square site through the photonic lattice. We repeat this recipe until the emitters site reaches the lattice boundary.

\section{Overlap between BIC photon and an input photon of arbitrary shape}\label{Appendix_sec_3}

Here we evaluate the overlap between the initial state $\ket{g,g}\otimes\ket{\varphi_{\pm}}\equiv\ket{gg,\varphi_{\pm}}$ and the BIC in Eq.~\eqref{eq:BIC_m}. We consider $\ket{\varphi_{\pm}}$ being a resonant input photon of the form $\ket{\varphi_{\pm}} = ( \ket{\varphi_R} \pm \ket{\varphi_L} ) / \sqrt{2}$ and arbitrary envelope: 
\begin{align}\label{eq_photon_input}
    \ket{\varphi_R} & = \int dx~\Theta(d-x) \xi(d-x) e^{-i k_0 (d-x)}c^{\dagger}_{R}(x) \ket{0}, \\ \nonumber
    \ket{\varphi_L} & = \int dx~\Theta(x) \xi(x) e^{-i k_0 x} c^{\dagger}_{L}(x) \ket{0},
\end{align}
 with $\int dx |\xi(x)|^2=1$. The overlap reads 
\begin{align} 
\bra{\Phi_{\pm}}gg,\varphi_{\pm}\rangle=\sqrt{\frac{\gamma}{2 v_g}} \left( 1 + \frac{\gamma\tau}{2} \right)^{-\frac{1}{2}}\bra{1_{\mathrm{BIC}}}\varphi_{\pm}\rangle , 
\end{align}
with
\begin{align}
\bra{1_{\mathrm{BIC}}}\varphi_{\pm}\rangle &=\frac{- i}{2} \int_{0}^{d} dx \left[ e^{-i k_0 d} \xi(d-x)  \mp \xi(x) \right]\\ \nonumber
&=\frac{\pm i}{2} \int_{0}^{d} dx \left[ \xi(d-x)  + \xi(x) \right],
\end{align} 
where in the last equality we used the fact that our protocol requires the symmetry-matching: $\ket{\varphi_{\pm}}=\ket{\varphi_{(-)^{n+1}}}$ with $n=k_0 d/\pi$. As the dynamics takes place in the single-excitation sector, then the BIC generation probability $P_{\mathrm{BIC}}$, in the case of ideal interaction switch, equals the modulus square of the above overlap.

If the input photon is a decreasing exponential of envelope $\xi(x') =\sqrt{\frac{\Gamma}{v_g}} e^{-\frac{\Gamma}{2v_g}x'}$, the probability of populating the BIC reads
\begin{align}
    P_{\mathrm{BIC}} & = \left| \bra{\Phi_{\pm}}gg,\varphi_{\pm}\rangle \right|^2 \nonumber \\ & =\frac{\gamma\tau}{(2+\gamma\tau)4d}\left\vert \sqrt{\frac{\Gamma}{v_g}}\int_{0}^{d} dx \left[ e^{-\frac{\Gamma}{2v_g}(d-x)}  + e^{-\frac{\Gamma}{2v_g}x} \right]\right \vert^2,
\end{align}
which upon integration gives Eq.~\eqref{eq:th_upp_bound}.

Notice that, the overlap with an input photon of opposite symmetry would have vanished. Using the above results, it is also straightforward to verify that $\left| \bra{\Phi_{\pm}}gg,\varphi_{R}\rangle \right|^2 =\left| \bra{\Phi_{\pm}}gg,\varphi_{L}\rangle \right|^2 =P_{\mathrm{BIC}}/2$.

\subsection{Weak coherent input}
Here we consider the input being an attenuated coherent field. For simplicity we take the coherent field being the product of two counterpropagating square pulses of duration $\tau$ and same intensity, whose relative phase matches the BIC symmetry: $\ket{\alpha_{R},\alpha_{L}}$ with $\alpha_{L}=\alpha/\sqrt{2}=(-)^{n+1}\alpha_{R}$ and $n=k_0 d/\pi$. If the field is weak enough, $|\alpha|^2\ll 1$, we can neglect components with more than one excitation and the dynamics is still confined in the single-excitation sector. This in turn implies that the probability of populating the BIC is still (approximately) equal to the overlap modulus square:
\begin{align}
P_{\mathrm{BIC}}\approx \frac{\gamma\tau}{(2+\gamma\tau)d}\left\vert \bra{1_{\mathrm{BIC}}}\alpha_{R},\alpha_{L}\rangle \right\vert^2
\end{align}
with
\begin{align}
\bra{1_{\mathrm{BIC}}}\alpha_{R},\alpha_{L}\rangle &=\biggl\langle 1_{\mathrm{BIC}} \biggr| \frac{\alpha}{\sqrt{2}},\frac{(-)^{n+1}\alpha}{\sqrt{2}}\biggr\rangle \\ \nonumber
&= \frac{- i}{2}
\left[\int_{0}^{d} dx \frac{\left( e^{-i n\pi} \alpha  + (-)^{n} \alpha \right)}{\sqrt{d}}\right] \bra{0}\alpha_{R},\alpha_{L}\rangle \\ \nonumber
&= i (-)^{n+1} e^{-|\alpha|^2/2} \alpha \sqrt{d};
\end{align}
giving $P_{\mathrm{BIC}}\approx\gamma\tau |\alpha|^2 e^{-|\alpha|^2}/(2+\gamma\tau)$, where the factor $ |\alpha|^2 e^{-|\alpha|^2}$ is upper-bounded by $0.37$, then, a comparison with Eq.~\eqref{eq:BIC_m} shows that an attenuated coherent field is less efficient than the single photon of optimal bandwidth $\Gamma=2.51/\tau$.  

%\bibliography{biblio}% Produces the bibliography via BibTeX.

%

\end{document}